# Ultrafast All-optical Modulation Exploiting the Vibrational Dynamic of Metallic Meta-atoms

Biqin Dong[1,2,*], Xiangfan Chen[1,*], Fan Zhou[1], Chen Wang[1], Hao F. Zhang[2], and Cheng Sun[1,†]

[1]Mechanical Engineering Department, Northwestern University, Evanston, IL 60208, USA

[2]Biomedical Engineering Department, Northwestern University, Evanston, IL 60208, USA

*These authors contributed equally to this work

† To whom correspondence should be addressed. E-mail: c-sun@northwestern.edu

**Optical control over elementary molecular vibration establishes fundamental capabilities for exploiting the broad range of optical linear and nonlinear phenomena[1-4]. However, experimental demonstration of the coherently driven molecular vibration remains a challenge task due to the weak optical force imposed on natural materials. Here we report the design of "meta-atom" that exhibits giant artificial optical nonlinearity. These "meta-atoms" support co-localized magnetic resonance at optical frequency and vibration resonance at GHz frequency with a deep-sub-diffraction-limit spatial confinement (~$\lambda^2$/100). The coherent coupling of those two distinct resonances manifests a strong optical force, which is fundamentally different from the commonly studied form of radiation forces[5], the gradient forces[5,6], or photo-thermal induced deformation[7]. It results in a giant third-order susceptibility $\chi^{(3)}$ of $10^{-13}$ m$^2$/V$^2$, which is more than six orders of magnitude larger than that found in natural materials[8]. The all-optical modulation at the frequency well above 1 GHz has thus been demonstrated experimentally.**

Metamaterials or metasurfaces are the assemblies of rationally designed sub-wavelength-scale "Meta-atoms" that feature non-natural occurring properties. Their unique ability to manipulate electromagnetic waves[9-12] have therefore inspired many fascinating applications, such as sub-diffraction limited imaging[13,14], invisibility cloaks[15-17], and near-zero-index materials[18-20]. It is worthwhile to note that their extraordinary properties are engineered through structures rather than through their constituting materials and thus, not only offers an extremely flexible platform for active control of material properties such as tunability, switching and modulation of electromagnetic waves[21-25], but also shed the light to facilitating strong optomechanical coupling[26-29]. Capturing upon the intriguing abilities for controlling electromagnetic waves, the coherently driven vibrational motion of meta-atoms with tailored mechanical properties has not been studied, which sets the foundation of all-optical switching in this study.

The meta-atom studied in this work comprises metallic nano-wire featuring a vertically orientated, U-shaped cross-section that supports co-localized optical and mechanical resonances (Fig. 1a). From a mechanical resonance perspective, the U-shaped structure bears characteristics similar to that of the molecule or the tuning fork[30], as its two freestanding prongs are free to vibrate. The nano-scaled tuning fork exhibits the lowest fundamental anti-symmetric mode (Fig. 1d) and symmetric mode (Fig. 1e) at the GHz frequency range; however, from an optical resonance perspective the U-shaped cross-section closely resembles the characteristic response of the split-ring resonator[10,11]. The U-shaped resonator has been widely used as the primary building block of metamaterials with tailored magnetic response, which is achieved by mimicking the LC resonance among a capacitor with capacitance $C$ and a magnetic coil with inductance $L$ (Fig. 1b). The resonance-induced electric currents circulating along the U-shaped cross-section not only guide the induced magnetic field within an extremely confined space far beyond the diffraction limit of the light (Fig. 1c), but also facilitate a strong repulsive force among the anti-parallel current along the freestanding prongs. The induced repulsive forces excite the symmetric vibration mode, and the resulting coherent motion of the two freestanding prongs modulates the capacitance *C* and inductance *L* and subsequently the effective refractive index due to the resulting changes in the underlying LC resonance. Unlike the commonly used optical forces, in the form of radiation forces and the gradient forces[5,6], or photo-thermal induced deformation[7], the U-shaped nano-wire array studied here provides a new way to facilitate the highly efficient opto-mechanical coupling by exploiting the strong electric force interaction and the superposition of optical and mechanical energy in a deep sub-diffraction-limited volume ($\sim\lambda^2/100$).

The nanotransfer printing (nTP) process[31] was used to fabricate the metasurface consisting a large array of U-shaped Au nanowires (meta-atoms) covering an area of $25\times25$ mm$^2$ on the glass substrate (Fig. 2a, see detailed process flow in Supplementary Information). A magnified view is shown in Fig. 2b, in which each U-shaped nanowire has a base width $w$=160 nm and thickness $b$=40 nm, and two freestanding prongs with height $h$=75 nm and thickness $t$=35 nm.

To better understand the optical resonance modes of the metasurface, numerical simulation using a commercial finite-difference time-domain solver (FDTD Solutions, Lumerical) was performed. Experimental data of the complex permittivity of gold was used in the simulations[32]. For convenience, an electric field perpendicular and parallel to the length of the U-shaped nanowire is referred as TE polarization and TM polarization, respectively. The LC resonance can be excited under the TE polarized illumination when the magnetic field vector has a component perpendicular to the U-shaped cross-section[33,34]. Fig. 2c shows the simulated transmission spectra for both TE and TM polarized light. As expected, the magnetic resonance can be found under the TE polarized illumination at the center wavelength of 815 nm while the commonly known electric quadrupole resonance can also be found near 530 nm[33,34]. The simulated magnetic field distribution at the corresponding magnetic resonance condition (λ=815 nm) suggests a strongly confined magnetic field within the cross-section area of 90 nm by 75 nm (Fig. 1c). In comparison, both magnetic resonance and electric quadrupole resonances do not exist under the TM polarized illumination.

The optical transmission spectra of the fabricated metasurface are measured experimentally (Fig. 2d). Two distinct resonances match closely with the expected electric quadrupole and magnetic resonance of the U-shaped resonator. In particular, a strong modulation with near-zero transmission is found at the long-wavelength resonance centered at 805 nm, which is in a good agreement with the theory. As expected, both resonances diminish under the TM polarized illumination.

The observed magnetic resonance induces the alternating electric current along the U-shaped cross-section and the charge accumulation at the tip of the prongs, both of which contribute to the direct electric forces acting upon the two freestanding prongs. However, our calculation suggests that the repulsive force among the anti-parallel current is dominant compared with the significantly weaker attractive Coulomb force among the opposite charges (see Supplementary Information). The repulsive forces exerted by the neighboring nano-wires are considerably weaker due to the relatively large spacing in between, resulting in the net repulsive force within the U-shaped nanowires. To understand the vibrational characteristics of the nano-scale tuning fork driven by the repulsive electric force, a modal analysis was performed using the finite element method (COMSOL Multiphysics). The lowest harmonic mechanical resonance mode consists of two slightly degenerated Eigen modes[30], which corresponds with the anti-symmetric and symmetric vibration of the two prongs near 1.5934 GHz (Fig. 1d) and 1.5938 GHz (Fig. 1e), respectively (see Supplementary Information). Considering the symmetry of the repulsive force originated from the optical resonance, only the symmetric vibrational mode can be selectively excited.

The magnetic resonance mode shifted as induced vibrations modulated the capacitance $C$ at GHz frequency. Since the optical field oscillates six orders of magnitude faster that the mechanic vibration at GHz frequencies, the optical transmission follows adiabatically to the vibration-induced shift in the magnetic resonance. The FDTD simulation was used to evaluate the coupled structural-optical effects. The vibration-induced displacement ($d$) of the freestanding prongs is illustrated as the inset in Fig. 3a. As $d$ changed from -10 nm to 10 nm, the magnetic resonance blue-shifted from 835 nm to 775 nm (Fig. 3a); the shift of the electric resonance mode was negligible. The corresponding shift in the magnetic resonance frequency $f_c$ is plotted in Fig. 3b. Moreover, the transmission at the probing wavelength of 780 nm is expected to decrease monotonously from 0.17 to close to zero (Fig. 3c). The modulation depth ($\Delta T/T$) near the equilibrium position shows pronounced wavelength dependent behavior (Fig. 3d). The maximum modulation depth near the magnetic resonance at 785 nm reaches 1.16 when the prong vibrates at the amplitude ($\Delta d$) of 1 nm. Thus, once excited, the symmetric vibrational motion can be used for dynamical modulation of the light at GHz frequencies due to the optical force induced effective refractive index change, which is similar to the optical Kerr effect[8].

A time-resolved optical pump-probe technique was used to investigate the artificial nonlinearity the metasurface (Fig. 1a; also see Supplementary Information). An ultra-short pulse laser (pulse duration of 120 fs and pulse energy of 25 μJ) was used as the pump source to excite the coherent vibration of the U-shape resonator. The pump beam was collimated with a Gaussian diameter of 1 mm. The wavelength of the pump laser was tuned to 790 nm (Pump I in Fig. 3a) to excite the magnetic resonance of the metasurface. The ultrafast excitation using the femtosecond laser induces an impulsive response that corresponds with an extremely wide frequency component and can excite the harmonic vibrational modes. A continuous wave diode laser (780 nm, 20 mW) was used to probe the induced optical modulation near the magnetic resonance (Fig. 3a). The time-resolved optical transmission of the probe beam was measured using a high-speed avalanche detector (APD210, Menlosystem). The time-resolved optical transmission from the metasurface was measured experimentally while the transmission from the epoxy-coated glass substrate is also measured as the control case (Supplementary Fig. 5). The measurement from the glass substrate shows a strong frequency response below 1.3 GHz, likely due to the thermal-induced vibrational resonance of the thin epoxy layer. In contrast, the measurement from the metasurface exhibits a well-defined peak near 1.85 GHz in addition to the features found in the control case, and thus a high pass filter with the cut-off frequency of 1.3 GHz was applied to remove signal from the background. The resulting transient optical transmission of the probe beam is shown as the red curve in Fig. 4a. The rapid oscillatory response is due to the ringdown of the mechanical vibration being excited by the ultra-fast pump laser. The maximum displacement of the prongs derived from the modulation depth of transient optical transmission is estimated to be around 0.164 nm near the equilibrium position. As shown in the inset of Fig. 4a, the modulation depth due to the artificial optical Kerr effect is proportional to the pump power. Since the modulation caused by system absorption at pump wavelength is negligible (see Supplementary Fig. 6), the real part of nonlinear third-order susceptibility $\chi^{(3)}$ of the metasurface is estimated to be at the order of magnitude of $10^{-13}$ $m^2/V^2$ (see details in Supplementary

Information), which is more than six orders of magnitude larger than that found in natural materials (typical $\chi^{(3)} \sim 10^{-19}$ -$10^{-22}$ $m^2/V^2$)[8].

The corresponding Fourier transform reveals a pronounced spectral peak at the center frequency of 1.85 GHz, which agrees reasonably well with the calculated resonant frequency of the fundamental symmetric vibration mode (Fig. 4b). The higher resonant frequency observed experimentally is likely caused by the rounded corner of the fabricated structure, which reduces the effective length of the freestanding prongs. Besides, the overall modulation is also affected by the inhomogeneous broadening in frequency domain due to manufacturing defects and structural variations across the metasurface.

To better understand the physical origin of the observed all-optical modulation, the control experiment was performed by tuning the wavelength of the probe beam to 650 nm to avoid excitation of the magnetic resonance (Pump II in Fig. 3a). The time-domain and frequency-domain response are plotted as the green curves in Fig. 4a and Fig. 4b, respectively. While the laser intensity was kept the same as the previous case, the characteristics associated with the harmonic vibration completely disappeared; therefore, the contribution from the direct radiation force can be eliminated because the resulting moment transfer from the reflecting photons can be far less efficient compared with the magnetic-resonance-mediated, strong opto-mechanical coupling reported here. Furthermore, the U-shape resonator exhibits similar absorption at 650 nm and 790 nm (see Supplementary Fig. 6); thus, the diminishing characteristic vibrational signature further excludes the possible contribution from photo-thermal effect.

An additional control experiment was performed using an Au grating that featured the same periodicity of the U-shaped nano-wire array (blue curves, Fig. 4). As expected, the vibrational characteristic diminished due to the lack of both the magnetic and mechanical resonance. Based on these findings, we can conclude that superposition of the magnetic and mechanical resonance in a deep-subwavelength volume plays the dominating role in facilitating the observed all-optical modulation using the metasurface, while the contributions from radiation force and the photo-thermal effect are negligible.

In conclusion, we presented here the design of a unique meta-atom comprising of U-shaped nano-wires with freestanding prongs that supports spatially overlapping magnetic resonance at optical frequency and vibration resonance at GHz. The magnetic-resonance-induced oscillating currents exert strong optical force to excite the vibration resonance of the freestanding prongs of the meta-atoms, which further modulates its effective optical properties. The coherent coupling of those two distinct resonance modes results in a giant third-order susceptibility six orders of magnitude larger than that found in natural materials. Such a pronounced optical nonlinearity further allows for the experimental demonstration of all-optical modulation of the transmitted light at 1.85 GHz. This work is not only promising for applications in optical isolation, filtering, and signal processing, it also sets the stage for exploiting the broad range of the non-linear optical phenomena and manifesting the quantum mechanical states by providing a completely new architecture for facilitating strong optical force interactions.

**Acknowledgements**

We also acknowledge the generous financial support from the National Science Foundation under Grant number CMMI-0955195, ECCS-1232134, DBI-1353952, CBET-1055379, and CBET-1066776.

## Figures

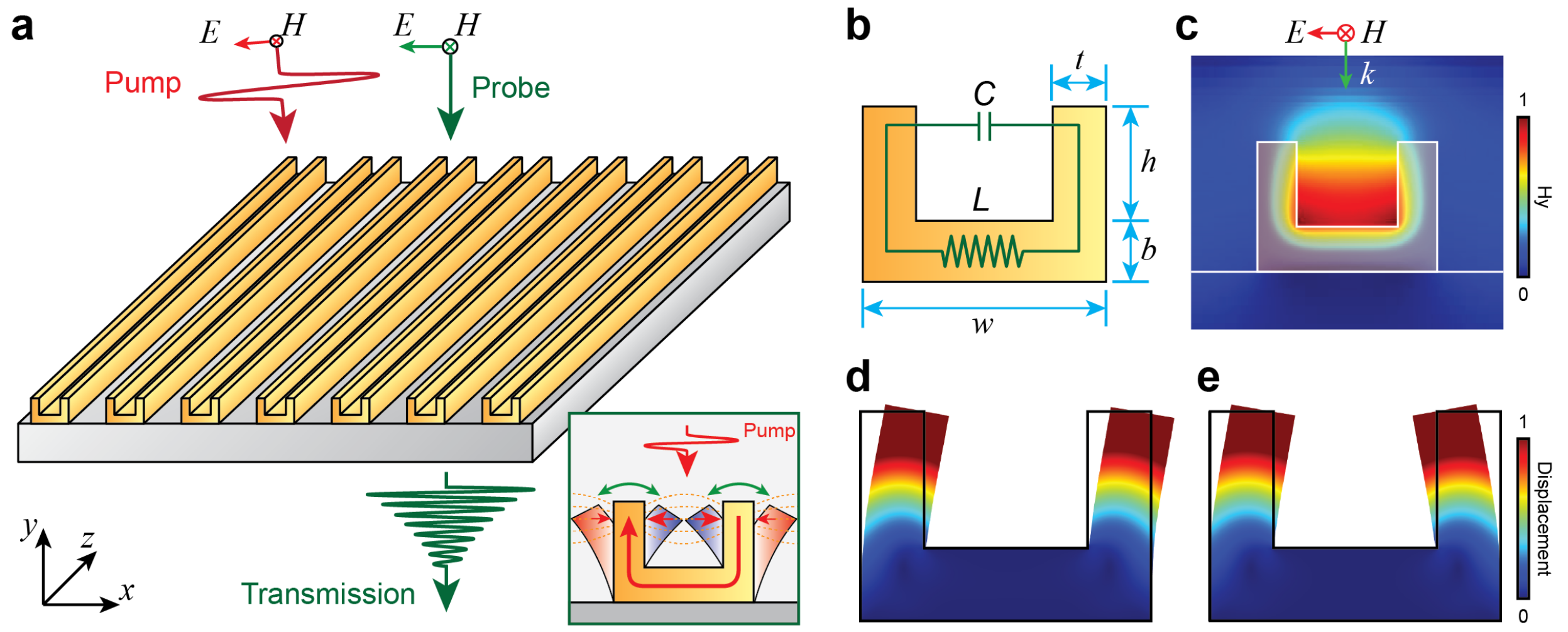

**Figure 1. Strong opto-mechanical coupling using metasurfaces exhibiting co-localized magnetic resonance at the optical frequency and mechanical resonance in the GHz region.** (a) Schematic drawing illustrates the all-optical switching using the pump-probe scheme. (b) Physical dimension of the U-shaped resonator and its equivalent LC circuit. (c) Simulated magnetic field along the U-shaped cross-section induced by TM incidence. Calculated mechanical motions of fundamental (d) anti-symmetric vibration mode at 1.5934 GHz and (e) symmetric vibration mode at 1.5938 GHz, respectively. The color map indicates the relative magnitude of the mechanical displacement.

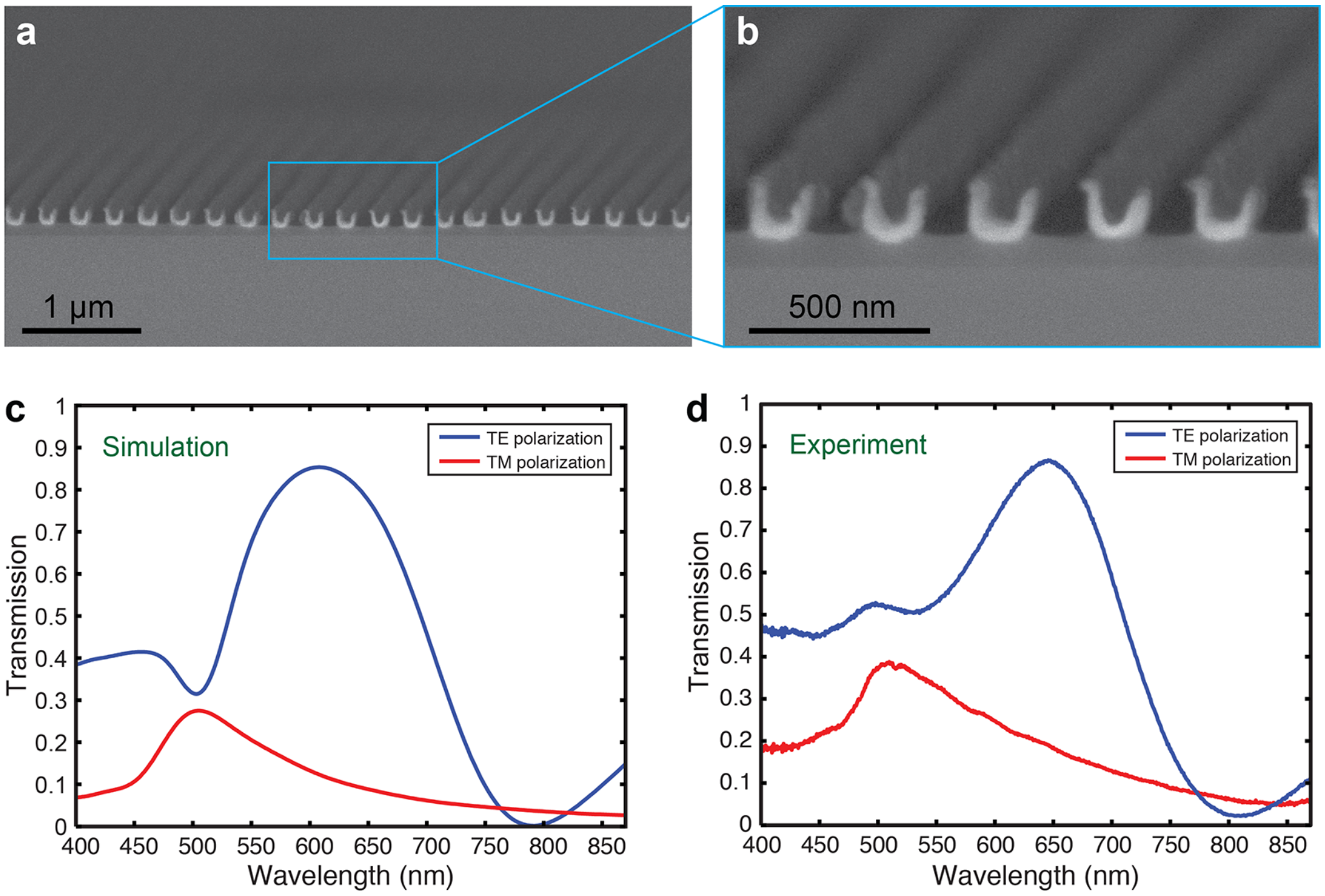

**Figure 2. Experimental measurements and numerical simulation for the optical response of the metasurface.** (a) Scanning electron micrograph of U-shaped Au nano-wires printed onto a glass substrate, and the magnified view (b). (c) The simulated transmission spectra by FDTD under TE and TM polarization, respectively. (d) The measured transmission spectra of the U-shaped resonators under TE and TM polarization, respectively.

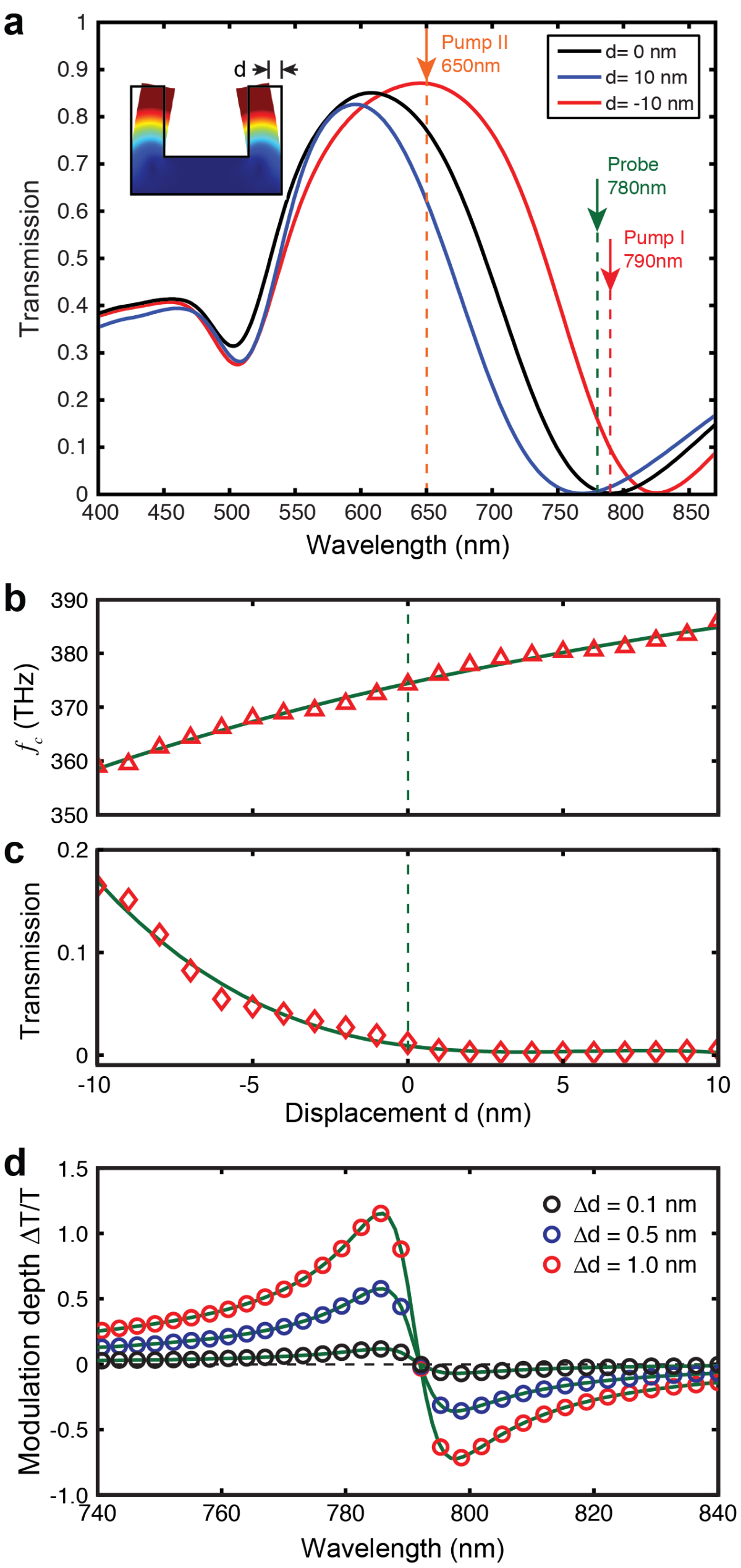

**Figure 3. Numerical simulation of vibration-induced light modulation.** (a) Optical transmission as a function of displacement *d* of the freestanding prongs. Inset shows the deformation of U-shaped resonator at symmetric vibration mode. Simulated resonant peak shift (b), transmission of the 780 nm probe beam (c) as the function of the displacement of the prongs. Solid lines are their polynomial fittings. (d) Modulation depth near the equilibrium position as the function of probing wavelength at different vibration amplitudes ($\Delta d$).

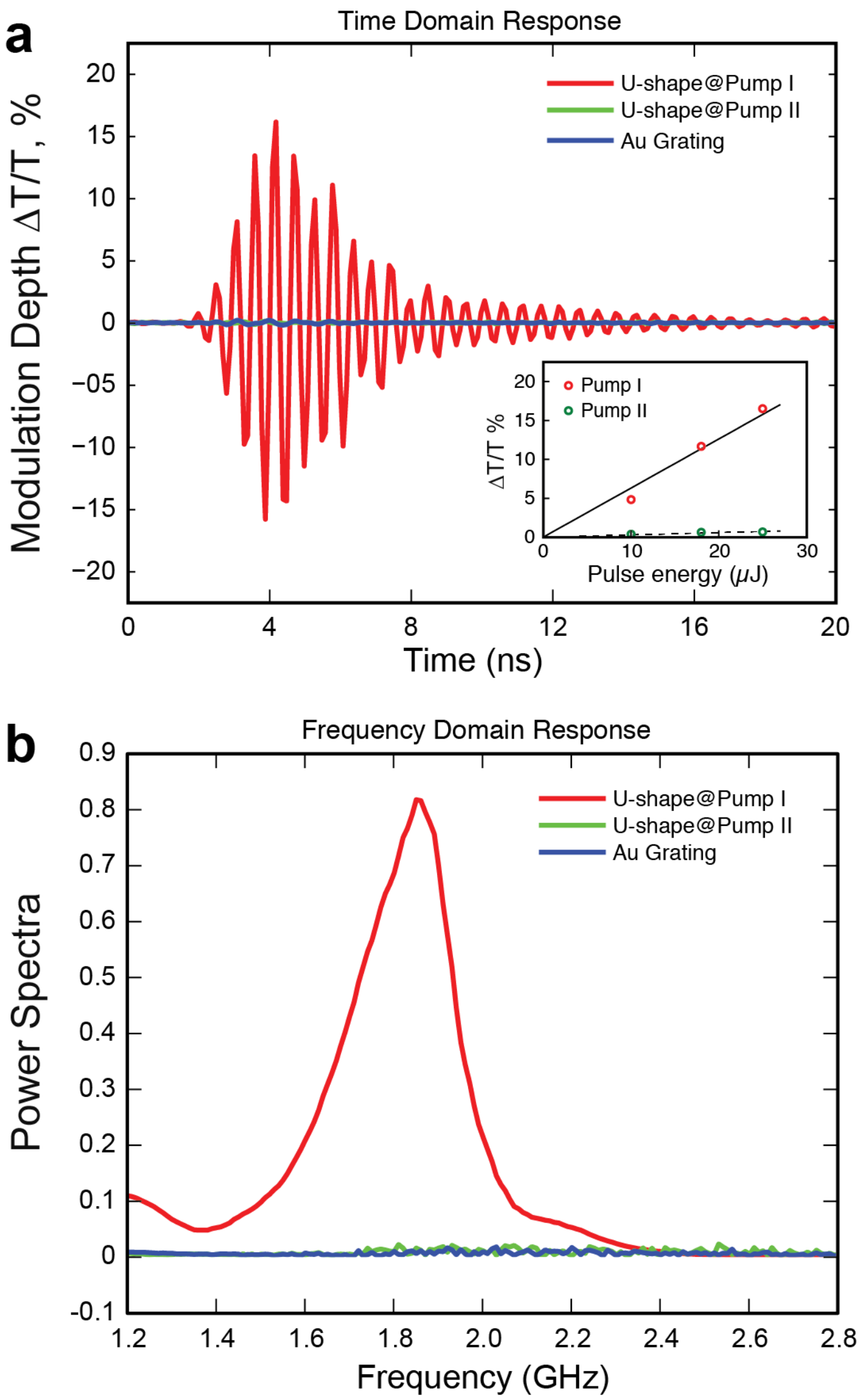

**Figure 4. Magnetic resonance-induced dynamic modulation of the U-shaped resonator.** (a) Time domain signal of experimentally measured optical demodulation of the induced vibration in U-shaped resonator and control samples. Inset: modulation depth as a function of pump laser power. (b) the corresponding Fourier transferred spectrum in frequency domain. Only the optical pump at the magnetic resonance (790 nm) exhibits significant modulation of the probe beam at 1.85 GHz, which is in good agreement with the modal analysis.

# Supplementary Information

# Ultrafast All-optical Modulation Exploiting the Vibrational Dynamic of Metallic Meta-atoms

Biqin Dong[1,2,]*, Xiangfan Chen[1,]*, Fan Zhou[1], Chen Wang[1], Hao F. Zhang[2], and Cheng Sun[1,†]

[1] Mechanical Engineering Department, Northwestern University, Evanston, IL 60208, USA

[2] Biomedical Engineering Department, Northwestern University, Evanston, IL 60208, USA

* These authors contributed equally to this work

† To whom correspondence should be addressed. E-mail: c-sun@northwestern.edu

## 1. Fabrication of Metallic Optomechanical Metamaterials

Metallic optomechanical metamaterials was fabricated using the Nanotransfer printing (nTP) process as shown in Fig. 1. The soft mold was made of UV-curable photopolymer that consists the mixture of 6 parts of poly[(mercaptopropyl)methulsiloxane] (PMMS, United Chemical Technologies), 4 parts of triallyl cyanurate (Sigma Aldrich), 1 part of ethoxylated bisphenol A dimethacrylate ester (Satomer, SR540), and 0.01 part of 2,2-dimethoxy-2-phenylacetophenone (Sigma Aldrich). The use of the soft mold makes it is easier to achieve close, conformal contact between the stamp and substrate and thus, increases the yield for success pattern transfer over larger area. The fabrication process starts with pre-fabricated silicon mold contains periodic one-dimensional (1D) gratings with the periodicity of 300 nm. The PMMS-based photopolymer is poured onto the silicon mold and the vacuum loading is used for degassing and improving the mold filling process. The PMMS-based photopolymer is then cured into the solid elastomer by exposing to the ultra-violate (UV) light for 5 minutes. It can then be easily peeled off from the silicon mold to create the soft stamp. The soft stamp was then coated with a thin layer of surfactant (Tridecafluoro-(1,1,2,2)-tetrahydrooctyl trichlorosilane, FOTS) as the releasing layer. The coating process is accomplished by immersing the stamp in a solution of surfactant dissolved in heptane. Then the directional depositions of gold are performed at approximately 45 degree from two opposite sides to form a U-shaped surface by using e-beam or thermal evaporation deposition method. The Au-coated stamp is then turned over and placed on a flat glass substrate with spin-coated 100 nm epoxy layer. The wetting between the U-shaped nanowire and the epoxy layer provides intimate contact in-between without the need for external pressure and thus, minimize the deformation of the soft mold. The epoxy layer can be cured by thermal activation or UV exposure. After curing process, the bonding of the metal to the epoxy is significantly strengthened. The soft stamp can then be peeled off from the substrate, leaving U-shaped nanowires array attached to the surface of the epoxy layer.

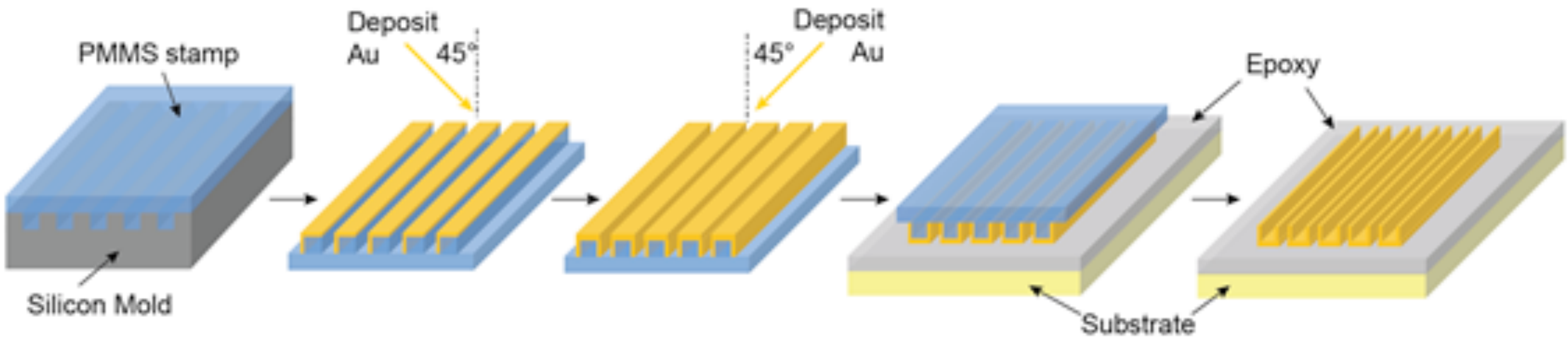

**Figure 1**. The nTP process flow of the U-shaped resonator array.

## 2. Estimation of Induced Electric Force

The magnetic resonance induces the alternating electric current along the U-shaped cross-section and the charge accumulation at the tip of the prongs, both of which contribute to the direct electric forces acting upon the two freestanding prongs. The magnitude of the induced electric forces due to the current-current interaction and charge-charge interaction are estimated using the using the average electric field strength ($E_{ave}$) within the gap opening and the average current density ( $J_{ave}$ ) moving along the prongs obtained from FDTD simulation. The incident electric field $E = 1\times 10^{9} Vm^{-1}$ is determined by the actual power density of the pump laser with pulse energy of 25 μJ, pulse duration of 120 fs, and beam diameter of 1 mm. As shown in Fig. 2, the electromagnetic field strength and current density are calculated to be $E_{ave} \approx 1.2\times 10^{9} Vm^{-1}$ and $J_{ave} \approx 1.0\times 10^{15} Am^{-2}$, respectively.

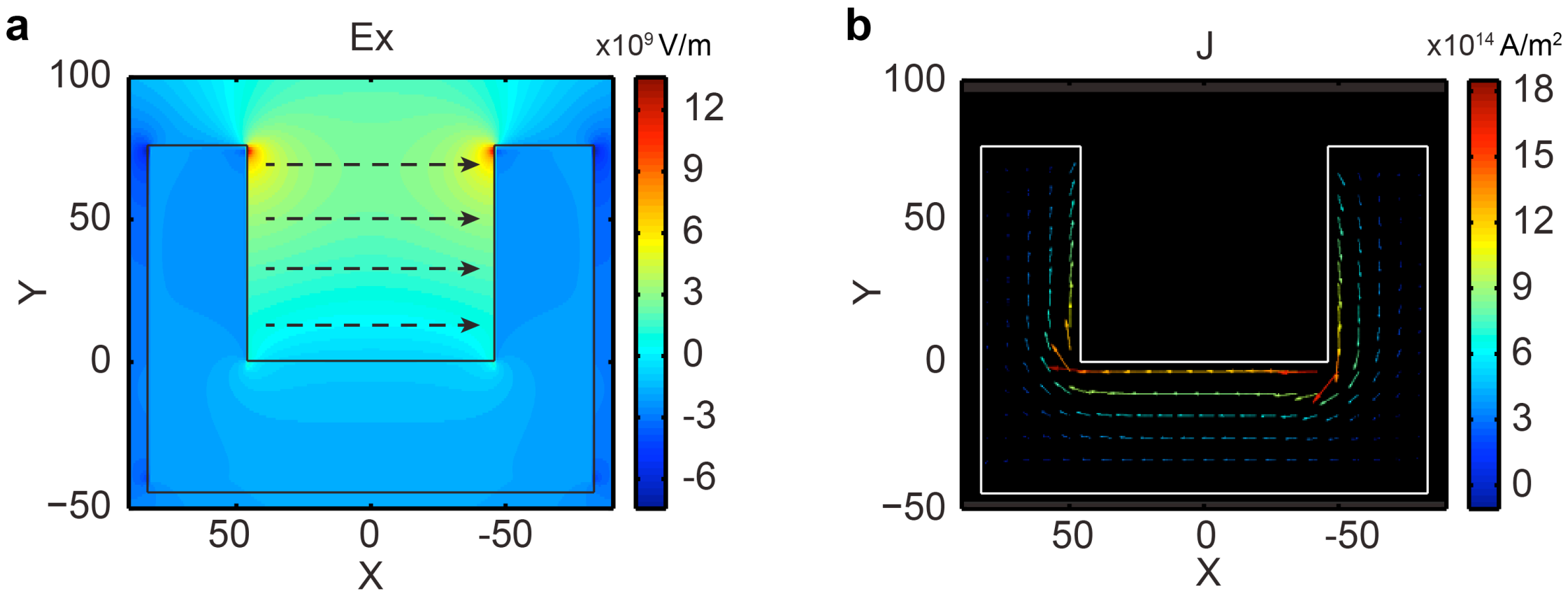

**Figure 2**. (a) Calculated electric field distribution (Ex) and (b) current density distribution (J) on *x-y* plane.

2.1. Charge-charge interaction

The two prongs are approximated as a pair of electrodes to form the parallel-plate capacitor. The accumulated charge on one of the prongs is approximately:

$$Q_{ave} = CE_{ave}d = \varepsilon_0 hlE_{ave}$$

where $\varepsilon_0$ is the electric constant. Thus, the electric force and pressure are given by:

$$F_{ee} = Q\cdot E$$
$$P_{ee} = \frac{F_{ee}}{hl} = \varepsilon_0 E_{ave}^2 = 1.275\times 10^{7} Nm^{-2}$$

2.2. Current-current interaction:

The induced current $I$ and magnetic field $B$ in a prong with a given length $l$ can be written as

$$I = tlJ_{ave}$$
$$B = \mu_0 I / 2l$$

where $\mu_0$ is the magnetic constant. According to the Lorentz force law, the current-current interaction induced force $F_{ii}$ and pressure $P_{ii}$ are given by:

$$F_{ii} = Ih \times B$$
$$P_{ii} = \frac{F_{ii}}{hl} = \frac{1}{2}\mu_0 t^2 J_{ave}^2 = 7.697 \times 10^8 Nm^{-2}$$

In the experiment, the thickness of the prongs is t=35 nm. Thus, the current-current interaction induced pressure is the dominating one while the contribution from the charge-charge interaction induced pressure can be negligible:

$$\frac{P_{ii}}{P_{ee}} = \frac{\mu_0 t^2 J_{ave}^2}{\varepsilon_0 E_{ave}^2} = 60.37$$

2.3. Optical force estimation using Maxwell stress tensor method

In addition, to verify the magnitude of the estimated induced electric force, Maxwell stress tensor method was also used to calculate the electric force between two prongs[1]. Maxwell stress tensor is the stress tensor of an electromagnetic field. As derived above in SI units, it is given by:

$$\sigma_{ij} = \varepsilon_0 E_i E_j + \frac{1}{\mu_0} B_i B_j - \frac{1}{2}(\varepsilon_0 \mathrm{E}^2 + \frac{1}{\mu_0}\mathrm{B}^2)\delta_{ij}$$

where *E* is the electric field, *B* is the magnetic field, and $\delta_{ij}$ is Kronecker's delta. Maxwell stress tensor was calculated from the electromagnetic field (*E* and *H*) obtained by FDTD simulation. In *x-y* plane, the total force on each prong is:

$$F_{MST} = 92.63 Nm^{-1}$$

Thus,

$$P_{MST} = \frac{F_{MST}}{h} = 1.235 \times 10^{9} \, Nm^{-2}$$

which is in a good agreement with the estimated current-current interaction induce force.

## 3. Modal Analysis of Vibration Modes

To investigate the vibrational characteristics of the U-shaped resonator, modal analysis was performed by using the finite element method (COMSOL Multiphysics 4.3a). Using eigenfrequency module and frequency domain module, we explored the vibrational response of the nano tuning fork under broadband excitation (0~6 GHz). In the simulation, the bulk properties of gold were chosen as the young's modulus is 79 GPa, the density is 19.3 g/cm$^3$ and passion ratio is 0.44 (from material library of COMSOL). The fundamental mechanical resonance mode consists of two slightly degenerated Eigen modes, $f_1$=1.5934 GHz and $f_2$=1.5938 GHz, which corresponds to symmetric and anti-symmetric vibration of the two prongs. Higher order harmonics around 5 GHz also exist but with much lower amplitude and faster decay in time according to the well-know mechanical properties of a conventional tuning fork[2].

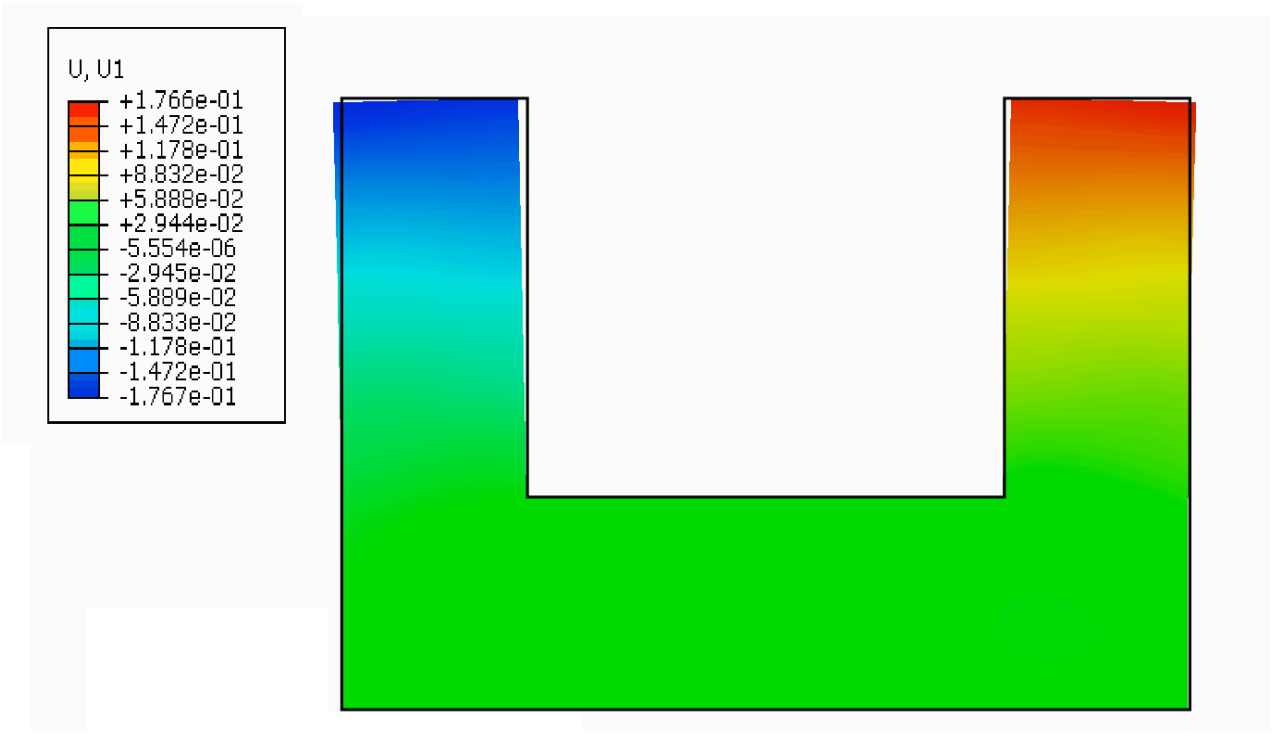

**Figure 3**. Maximum displacement of the prongs calculated by FEM method.

## 4. Estimation of the optical resonance induced displacement

To estimate the displacement of the prongs under the electromagnetic force, we assume that each prong of the fork experiences a force for a specific amount of time that results in a change in momentum. The result of the force acting for the given amount of time is that the object's mass either speeds up or slows down (or changes direction). The impulse experienced by the object equals the change in momentum of the object. In equation form:

$$J = \Delta p \Rightarrow \int F\,dt = m \cdot \Delta v$$

where $J$ is the impulse and $\Delta p$ is the change in linear momentum.

In addition, the bending process can be approximated as rotation process, so we can use the angular momentum theorem:

$$J = \Delta p \Rightarrow \int Fh\,dt = I \cdot \Delta w$$

$$I = \frac{1}{3}mh^2$$

So

$$P_{MST} Sh \Delta T = \frac{1}{3}mh^2 \omega$$

Then, using the free vibration model to calculate the maximum amplitude of the prongs' vibration:

The initial displacement of the prong is $x_0$, initial velocity is $\dot{x}_0$:

$$x_0 = 0, \dot{x}_0 \approx \omega h$$

And the effective equivalent stiffness $k$:

$$k = 3\frac{EI}{h^3} = \frac{1}{4}\frac{Elt^3}{h^3}$$

So the maximum displacement of the vibrating prong (amplitude of the vibration)

$$d_{\max} = A = \sqrt{x_0^2 + \dot{x}_0^2 \frac{m}{k}} = \omega h \sqrt{\frac{4\rho h^4}{Et^2}} = 0.2222nm$$

In addition, finite element method (FEM) was also used to calculate the maximum displacement of the prongs experiences the impulse, as shown in Fig. 3. The result suggested the maximum displacement is about 0.177 nm, which is in good agreement with the estimated results.

**5. Optical Kerr Effect**

The Kerr nonlinearity from the optical modulation presented by the metasurface can be quantified via the expression relating the refractive index with the intensity $I$ of the electromagnetic wave: $n = n_0 + n_2 I$, where $n_0$ are the linear refractive index[3]. The nonlinear refractive index $n_2$ can be estimated as $\Delta T/(It)$, where $\Delta T$=0.0045 excited by a 120 fs laser pulse with the pulse energy of 25μJ and $t$=115 nm is the thickness of the metasurface. The nonlinear refractive index is proportional to the real part of the nonlinear third-order susceptibility $\chi^{(3)}$ through the expression $\chi_R^{(3)} = (4/3)n_0^2 \epsilon_0 c n_2$, where $n_0$ and $\epsilon_0$ denote the linear refractive index of the material and the vacuum permittivity (8.85 x$10^{12}$ F/m), respectively. Thus the artificial $\chi^{(3)}$of the metasurface is estimated to be approximately 5x$10^{-13}$ $m^2/V^2$.

**6. Optical Measurement of All-Optical Ultrafast Modulation**

In experiment, a pump-probe method was used to test the vibrational properties of the U-shaped resonator, as shown in Fig. 4. A short-pulsed laser (Solstice, Spectra-physics) was used as pump source. The wavelength of pump source was further tuned to 650 nm and 790 nm by the optical parametric amplifier (OPA-800CF, Spectra-physics) with a harmonics module. The pulse duration is 120 fs and the repetition rate is 1 KHz. A variable ND filter is added to regulate the pump power. The collimated pump beam with Gaussian diameter of 1 mm was used to excited the vibration modes of samples at 15° incidence, while a continuous wave diode laser (780 nm, 20 mW) was used to probe the induced optical modulation. The polarization of pump and probe beams are both perpendicular to the elongated direction of U-shaped resonator. The transmitted light from the probe beam was filtered by a narrow band-pass filter (Ø1" Laser Line Filter, CWL = 780 ± 2 nm, FWHM = 10 ± 2 nm) to further reject the pump beam and then recorded by a high-speed avalanche detector (APD210, Menlosystem).

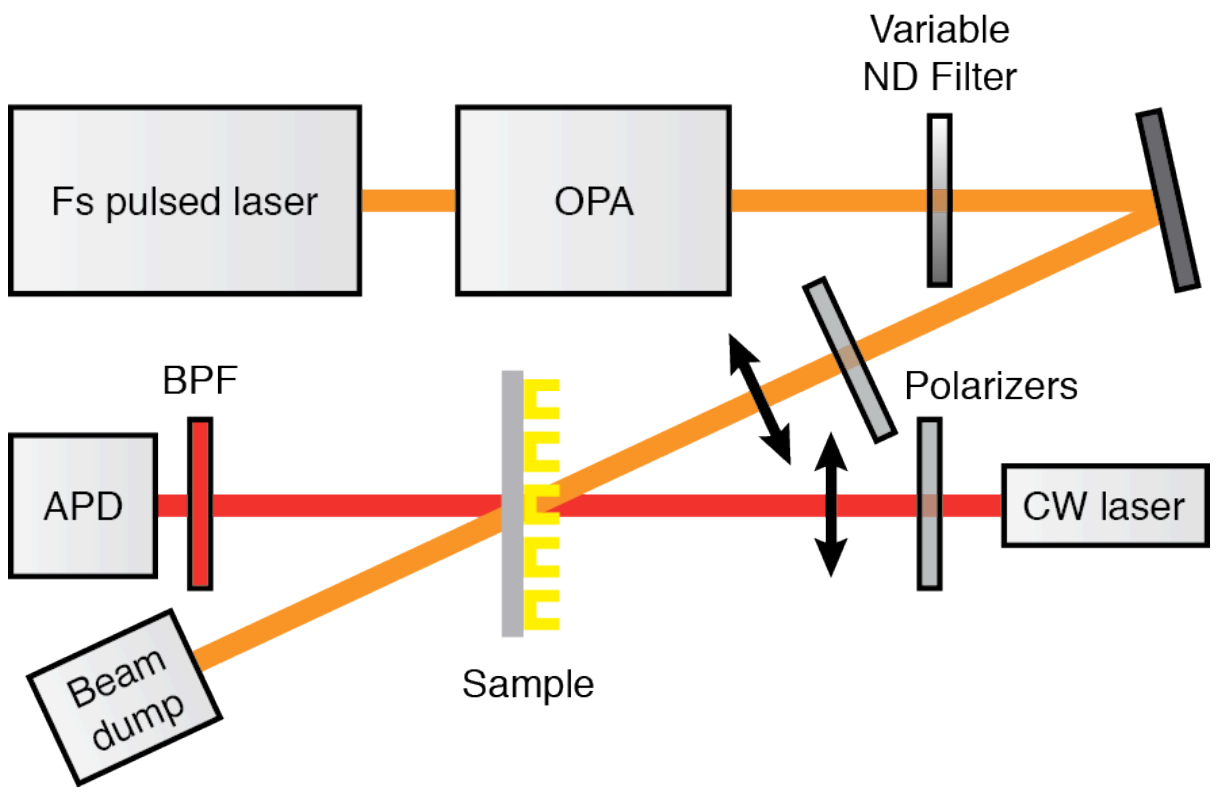

**Figure 4**. Schematic of experimental setup for measuring the dynamic modulation of U-shaped resonator.

The measured transient optical transmission of the probe beam for the U-shaped resonators sample and control sample are shown in Fig. 5. Fig. 5b shows the corresponding Fourier transform of the two samples. Multiple peaks below 1.3 GHz can be found for both of the samples, which can be considered as background, likely due to the thermal-induced vibrational resonance of the thin epoxy layer underneath the metasurface. In contrast, a pronounced spectral peak at the center frequency of 1.85 GHz is found to be unique to the U-shaped resonator, which agrees reasonably well with the calculated resonant frequency of the fundamental symmetric vibration mode.

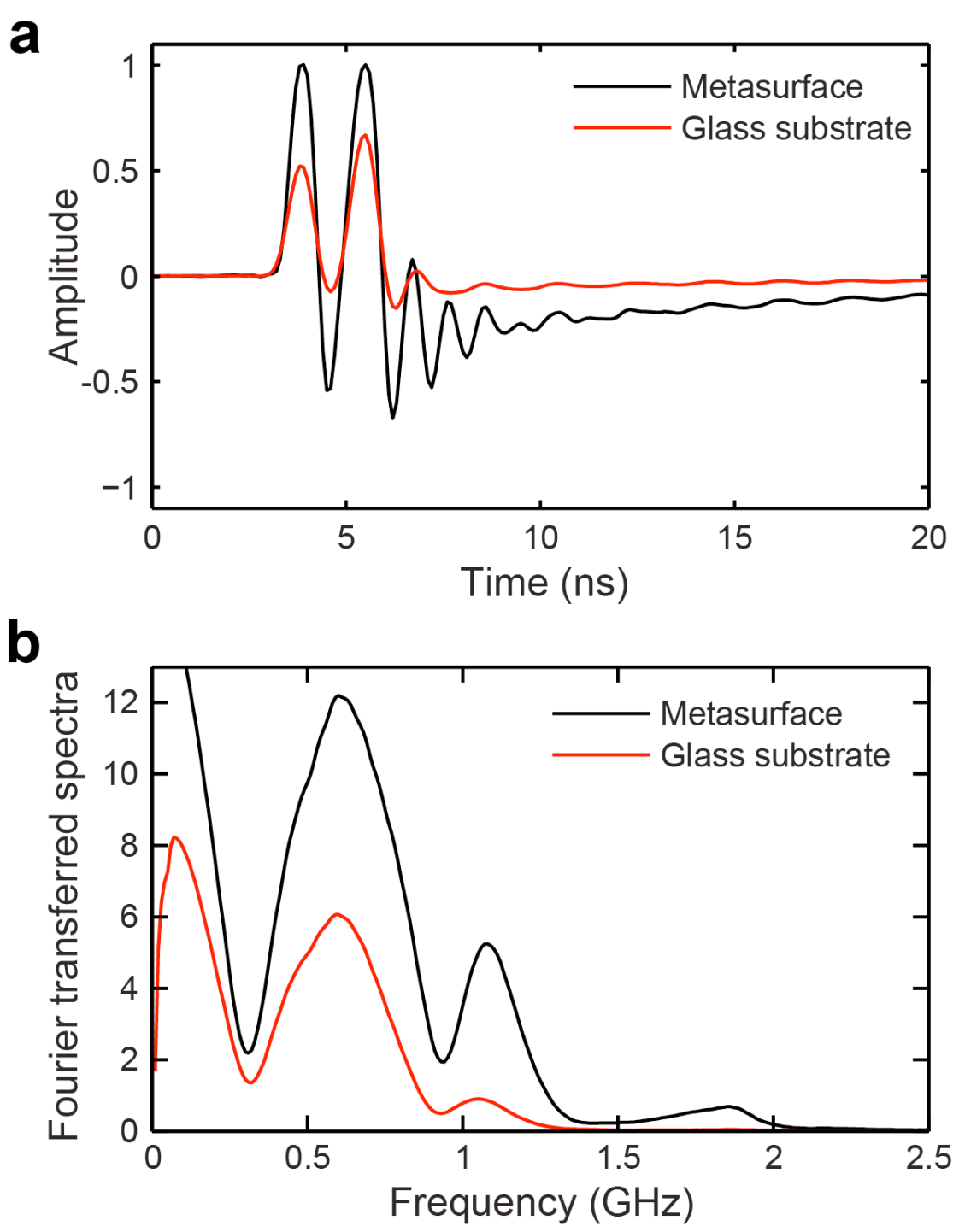

**Figure 5**. Magnetic resonance induced dynamic modulation of the U-shaped resonator. (A) Time domain signal of experimentally measured optical demodulation of the induced vibration in U-shaped resonator and the bare glass substrate as the control sample and (B) the corresponding spectra in frequency domain.

## 7. Absorption of the Metasurface

The absorption of the optomechanical metasurface under TE polarization was numerically calculated using a commercial finite-difference time-domain solver (FDTD Solutions, Lumerical). Experimental data of the complex permittivity of gold was used in the simulation[4]. To be noticed, the magnetic resonance at wavelength of 815 nm has much less absorption compared with the electric quadrupole resonance occurring at 530 nm. In order to verify the contribution of photo-thermal effect, 650 nm pump with comparable absorption was chosen for control experiment.

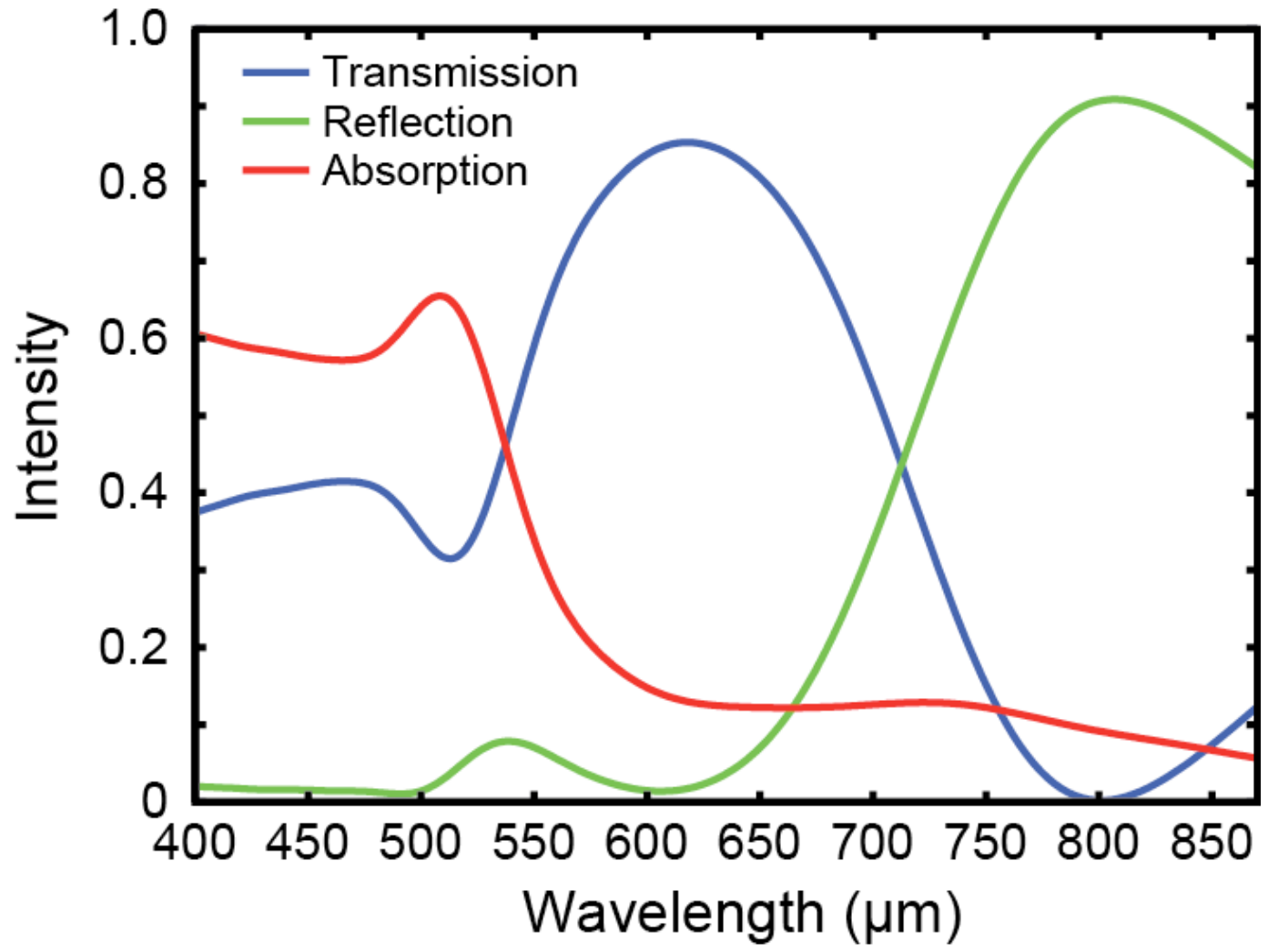

**Figure 6**. Simulated absorption of the metasurface under TE polarization.